# Sound Trapping in an Open Resonator

Lujun Huang[1#], Yan Kei Chiang[1#], Sibo Huang[2#], Chen Shen[3#], Fu Deng[1], Yi Cheng[2], Bin Jia[2], Yong Li[2*], David A Powell[1*] and Andrey E Miroshnichenko[1*]

[1] School of Engineering and Information Technology, University of New South Wales, Canberra, Northcott Drive, ACT, 2600, Australia

[2]Institute of Acoustics, Tongji University, Shanghai, 200092, People's Republic of China

[3]Department of Mechanical Engineering, Rowan University, Glassboro, NJ, 08028, USA

# These authors contributed equally to this work

[*] yongli@tongji.edu.cn, david.powell@adfa.edu.au, andrey.miroshnichenko@unsw.edu.au

**The ability of extreme sound energy confinement with high-quality factor (Q-factor) resonance is of vital importance for acoustic devices requiring high intensity and hypersensitivity in biological ultrasonics, enhanced collimated sound emission (i.e. sound laser) and high-resolution sensing. However, structures reported so far demonstrated a limited quality factor (Q-factor) of acoustic resonances, up to several tens in an open resonator. The emergence of bound states in the continuum (BIC) makes it possible to realize high-Q factor acoustic modes. Here, we report the theoretical design and experimental demonstration of acoustic BICs supported by a single open resonator. We predicted that such an open acoustic resonator could simultaneously support three types of BICs, including symmetry protected BIC, Friedrich-Wintgen BIC induced by mode interference, as well as a new kind of BIC: mirror-symmetry induced BIC. We also experimentally demonstrated the existence of all three types of BIC with Q-factor up to one order of magnitude greater than the highest Q-factor reported in an open resonator.**

# 1. Introduction

Acoustic resonators constitute the fundamental building block for acoustic metamaterials and metasurfaces[1–4]. They have been widely used in various design, such as acoustic absorbers and wavefront engineering[5,6]. Typically, most acoustic resonators reported so far have relatively low Q-factor limited to few tens, which may hinder their applications as acoustic sources (e.g. sound laser) and sensors. As a particular type of resonance, bound states in the continuum (BIC), also referred to as embedded trapped modes, have triggered extensive interest within the photonic community because they support zero radiative decay rate and infinite Q-factor[7–13], allowing for enhanced light-matter interaction[14–17]. In analogue with photonic BICs, acoustic systems such as a rigid plate in a waveguide and axisymmetric duck-cavity, also support BICs when losses are neglected[18–23]. However, all of these studies are theoretical works. More recently, the first experimental demonstration of quasi-BIC was done with a two-cavity system by breaking the symmetry[24], where the measured Q-factor was only around 50. In this work, we demonstrate that an open acoustic resonator could simultaneously support three types of BICs, including symmetry protected BICs (Fig.1a-b), mode interference induced BICs (Fig.1c-d), as well as newly observed mirror-symmetry induced BICs (Fig.1e-f). We also experimentally confirm the existence of these BICs. The measured Q-factors for these BICs are up to 250, 583, and 393, respectively. To the best of our knowledge, these are the largest Q-factors of acoustic resonances reported so far.

# 2. Results and discussion

We start by investigating the eigenmode properties of Helmholtz-like resonator system shown in Fig.2a. Such an open system can be treated as a closed rectangular acoustic cavity suject to deformation. A small air gap is opened to introduce the interaction between the cavity and the exterior environment. The role of the acoustic waveguide connecting to the neck is to guide the incoming acoustics waves, enabling measurement in experiments, as will be described in the later section. Because of its non-Hermitian nature, this open system turns the closed cavity modes into leaky modes (also denoted as quasinormal modes). The leaky modes have complex eigenfrequencies $\omega=\omega_0-i\gamma$, where $\omega_0$ and $\gamma$ are the resonant frequency and radiative decay rate, respectively. The radiative Q-factor can be derived from $Q=\omega_0/(2\gamma)$. Here, each leaky mode's complex eigenfrequency is calculated by COMSOL Multiphysics. Following the definition in Ref [20], the leaky modes are labelled as $M_{pq}$, where p and q are the number of maxima in the pressure field along the x- and y-axes, respectively.

We demonstrate that the reflection or transmission spectrum of such Helmholtz resonators can be perfectly reproduced with a complex eigenfrequency of leaky modes based on coupled-mode theory (CMT)[25] (See Section 1 and Fig.S1 in supporting material (SM)). Thus, the goal in searching BICs is to find leaky modes with infinitely large Q-factor. Since a BIC is also accompanied by vanishing linewidth of the acoustic resonance in the reflection or transmission spectrum, we calculate the reflection coefficient as a function of frequency and size ratios in Fig. S2 while assuming the system is lossless. Many BICs can be found by checking the vanishing linewidth of resonant peaks. These BICs can be categorized into three types (See Fig.1): symmetry protected BICs (Fig.1a-b), Friedrich-Wintgen BICs induced by two mode interference (Fig.1c-d) and mirror-effect induced BICs (Fig.1e-f). Except where explicitly mentioned, only the lossless case is studied for calculating the leaky modes. The realistic system's losses will deteriorate the total quality factor by $1/Q=1/Q_{abs}+1/Q_{rad}$ and will be discussed in the experiment section.

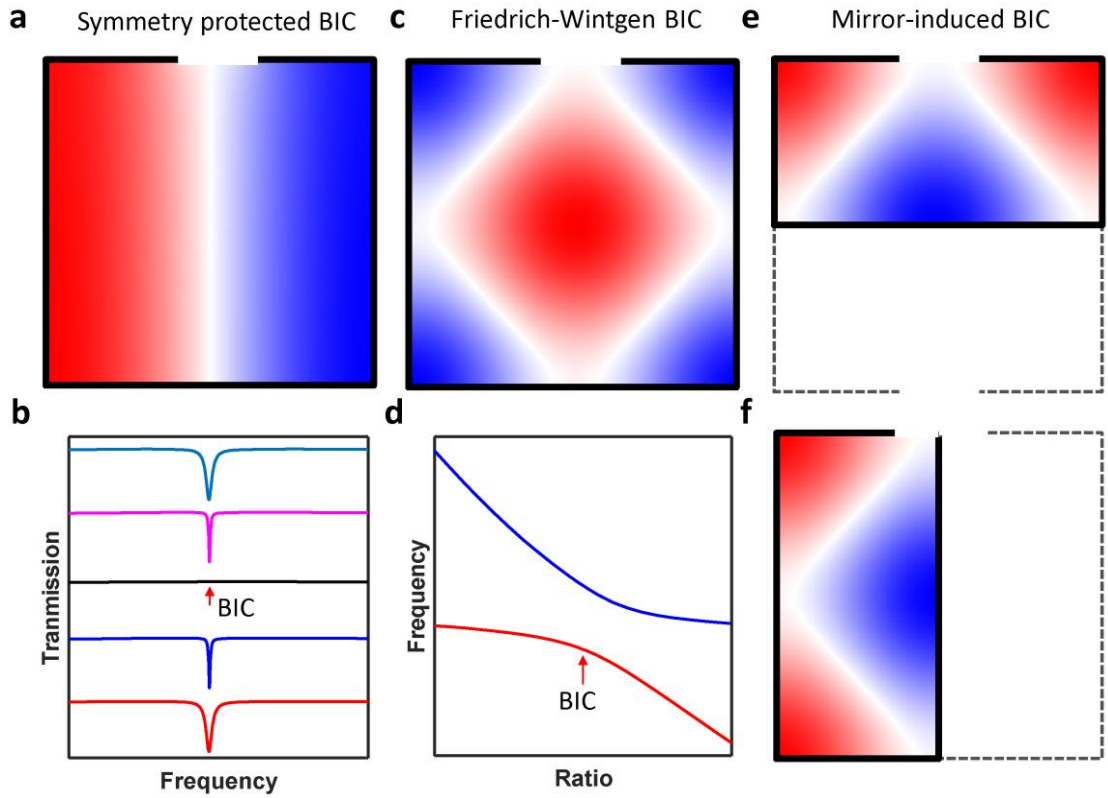

**Fig.1| BIC in an open acoustic resonator**. **a,** Symmetry-protected BIC. **b,** Transmission spectra evolution as structure asymmetry is introduced. When the asymmetry ratio decreases to 0 (black curve), ideal BIC occurs. **c,** Friedrich-Wintgen BIC induced by mode interferences. **d,** Resonant frequency of

high-Q and low-Q modes as a function of size ratio of the open resonator. **e,** Mirror-induced BIC for mirror along the x-axis. **f,** Mirror-induced BIC for mirror along the y-axis.

## 2.1 Symmetry-protected quasi-BIC

Referring to Fig. 2a, we consider a rectangular cavity of dimensions Lx by Ly. If only one open port is introduced on the left of the acoustic resonator, symmetry is broken along the y axis but still maintained along the x axis. Thus, it will support several symmetry protected BICs as long as the centre of neck and resonator are symmetric with respect to the x-axis. The left waveguide's existence will further lead to broken symmetry along the x-axis, and thus turn an ideal BIC into a quasi BIC. Without loss of generality, we use mode $M_{12}$ in the square cavity as an example to describe the effect of neck width w and centre shift $y_c$ on the Q-factor of BIC. Interestingly, from Fig.2b, the Q factor is still larger than $10^4$ for the system with protected symmetry along the y axis even when the neck's width is half of the right cavity width. The Q-factor increases exponentially with decreasing neck width. This can be understood by treating the neck as a perturbation of the square cavity. The smaller neck width, the smaller the perturbation, and thus the larger the Q-factor is. Another interesting finding is that the Q-factor is proportionally to $1/(y_c)^2$ when the neck width is small compared to the width of the right cavity, as shown in Fig.2c. This phenomenon has been observed and proved for BIC in photonic systems[26]. Additional symmetry protected quasi BICs can be found in such structures as long as q is even number for mode $M_{pq}$(See Fig.S3).

## 2.2 Friedrich-Wintgen BIC induced by mode interference.

Friedrich and Wintgen demonstrated that in quantum mechanics full destructive interferences of two degenerate modes gives rise to the avoided crossing of eigenvalues, accompanied by the formation of a BIC[27]. This type of BIC can be easily constructed in our system by tuning the size ratio of a rectangular cavity. For a closed rectangular cavity, cavity modes $M_{pq}$ and $M_{p+2,q-2}$ will become degenerate at a certain size ratio. When an air gap is introduced on one side of the rectangular cavity, strong coupling between these two modes results in the giant enhancement of Q-factor for one mode but suppresses it for another. We use paired modes $M_{23}$ and $M_{41}$ as an example to illustrate this principle. The pressure distributions for these two modes are shown in the inset of Fig.2d. It is found that the eigenfrequencies for these two modes cross at R=Lx/Ly=1.42 in a closed cavity. All the eigenmodes become leaky modes with complex eigenfrequencies when introducing the neck to couple the acoustic waveguides to the cavity. Based on the eigenmodes analysis, we find that the real part of eigenfrequencies exhibits an avoided crossing. Simultaneously, the Q-factor for mode $M_{23}$ is enhanced to a maximum of $6.87\times10^7$ but suppressed to minimum 86.76 for mode $M_{41}$ at R=1.398, as shown in Fig.2e-f.

Moreover, the avoided crossing suggests that these two modes interchange with each other after the size ratio passes through the critical size ratio. This interesting phenomenon is confirmed by the mode evolution, as shown in Fig.S4. Indeed, the upper branch mode, for example, evolves from mode $M_{23}$ into mode $M_{41}$ when the size ratio increases from 1.3 to 1.5. Following a similar strategy, more BIC induced by mode interference, such as $M_{13}$-$M_{33}$ and $M_{33}$-$M_{51}$, can be found by merely constructing avoided crossing (See Fig.S5a-d). Besides, we find that BIC can also be found in mode crossing for $M_{24}$ and $M_{42}$ (See Fig.S5e-f). Here, we emphasize that not all crossings of two modes of a closed cavity can bring about BICs. The two modes must have the same parity along both x- and y-directions. This is also the reason why we choose paired modes $M_{pq}$ and $M_{p+2q-2}$ here.

### 2.3 Mirror-induced BIC

In addition to the abovementioned two types of BICs that have been intensively studied in the photonics community, we also find a new type of BIC: mirror-symmetry induced BIC (See Fig.1e-f). Because all the outer boundary conditions are set as a hard wall in simulation, we can view the rightmost boundary as a partial mirror. All the eigenmodes with almost symmetric pressure distribution in the full-size resonator can also be found in a half-size resonator. Thus, many other BICs can also be constructed by simply shrinking the width to half based on this mirror effect. For example, one can easily find a BIC at R=0.498 for mode $M_{13}$ (Fig.2g), which is indeed half of the critical size ratio for BIC $M_{13}$ in a full resonator (Fig.2f).

Moreover, we find that the mirror effect also occurs for the x-axis. As shown in Fig.2i, a BIC can be found at R=1.986. Following a similar approach, more BICs can be constructed (See Figs.S6-7). The mirror effect indicates that one can achieve extreme pressure confinement even with reduced size in the 2D case, suggesting an effective way to engineer the Purcell factor that is the key to realize enhanced acoustic emission[28].

Note that the above three BIC types are not limited to regular rectangular shaped resonators. We can also find them in the elliptical resonator (See Figs.S8). The only difference is that the size ratio is defined as R=a/b, where a and b are semi-major and semi-minor axes, respectively. Besides, the conclusion drawn in the 2D case can be straightforwardly generalized to three-dimensional (3D) open resonators (e.g. cuboid resonators) (See Figs.S9-11). More freedom is provided in the 3D case because three parameters including length, width and height, are involved in the mode calculation.

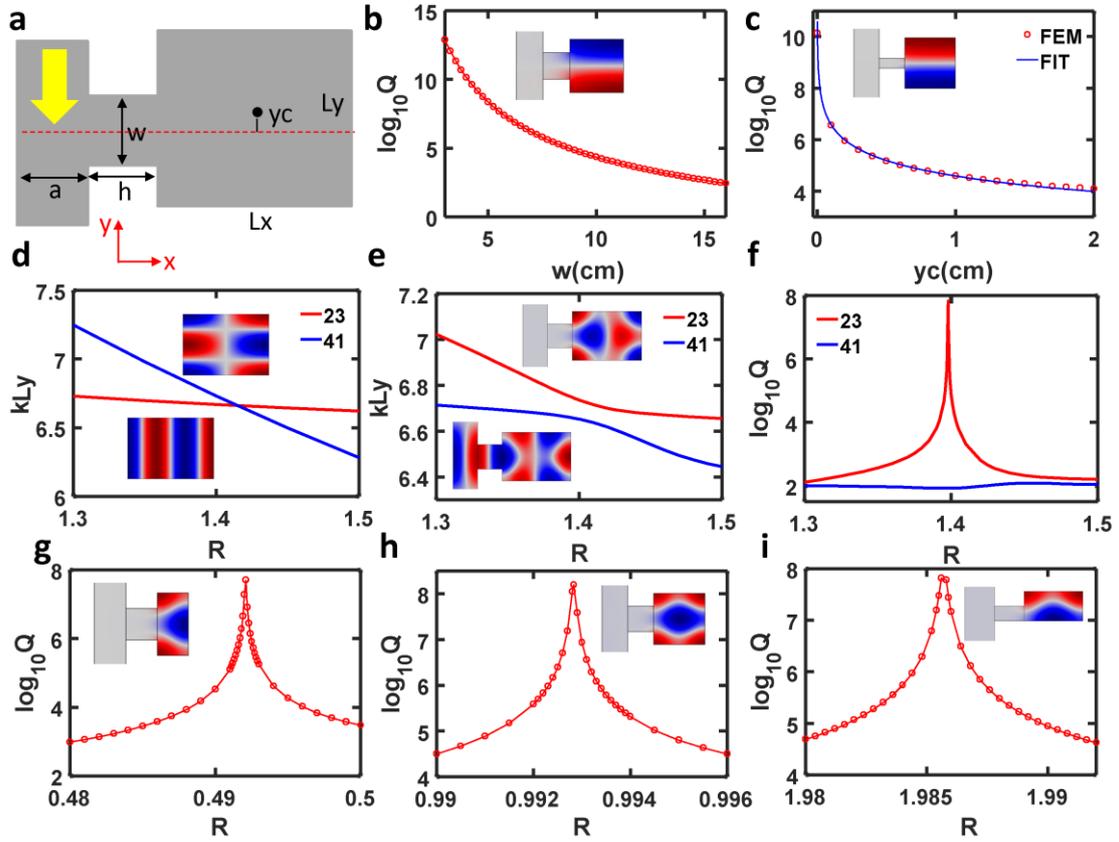

**Fig.2| BIC in an open acoustic resonator**. **a,** Schematic drawing of an open acoustic resonator, where an acoustic waveguide is coupled to a rectangular resonator. **b,** Q-factor of mode $M_{12}$ vs neck width. **c,** Q-factor of mode $M_{12}$ vs $y_c$. **d,** eigenfrequency kLy of mode $M_{41}$ and $M_{23}$ vs size ratio in a closed resonator. **e,** eigenfrequency kLy of mode $M_{41}$ and $M_{23}$ vs size ratio in an open resonator. **f,** Q-factor of mode $M_{41}$ and $M_{23}$ vs size ratio in an open resonator. **g,** Q-factor of mode $M_{13}$ in a half-open resonator. **h,** Q-factor of mode $M_{13}$ in a full resonator. **i,** Q-factor of mode $M_{13}$ in another half resonator.

## 2.4 Experimental verification of BIC

Next, we switch to the experimental demonstration of all three types of BICs. We fabricate two acoustic cuboid resonators shown in Fig.3a: full resonator and half resonator. In experiments, the left circular tube's diameter is fixed as d=29mm while the length, width, and height for the neck are set as 40mm, 20mm, 20mm, respectively. Fig.3b shows the measurement set up while Fig.4a-c depicts the structure's schematic. Fig.3c shows the measured transmission spectra for the full resonator and half resonator, and Fig.3d-g corresponds to the zoomed-in range in the vicinity of the BICs while Fig.3h shows the pressure distribution of BICs. Excellent agreement can be found between simulation and experiment over the full spectrum (See Fig.S12). Other leaky modes that are not BICs are given in Fig.S13. In the following, we discuss all three types of BIC observed in experiments.

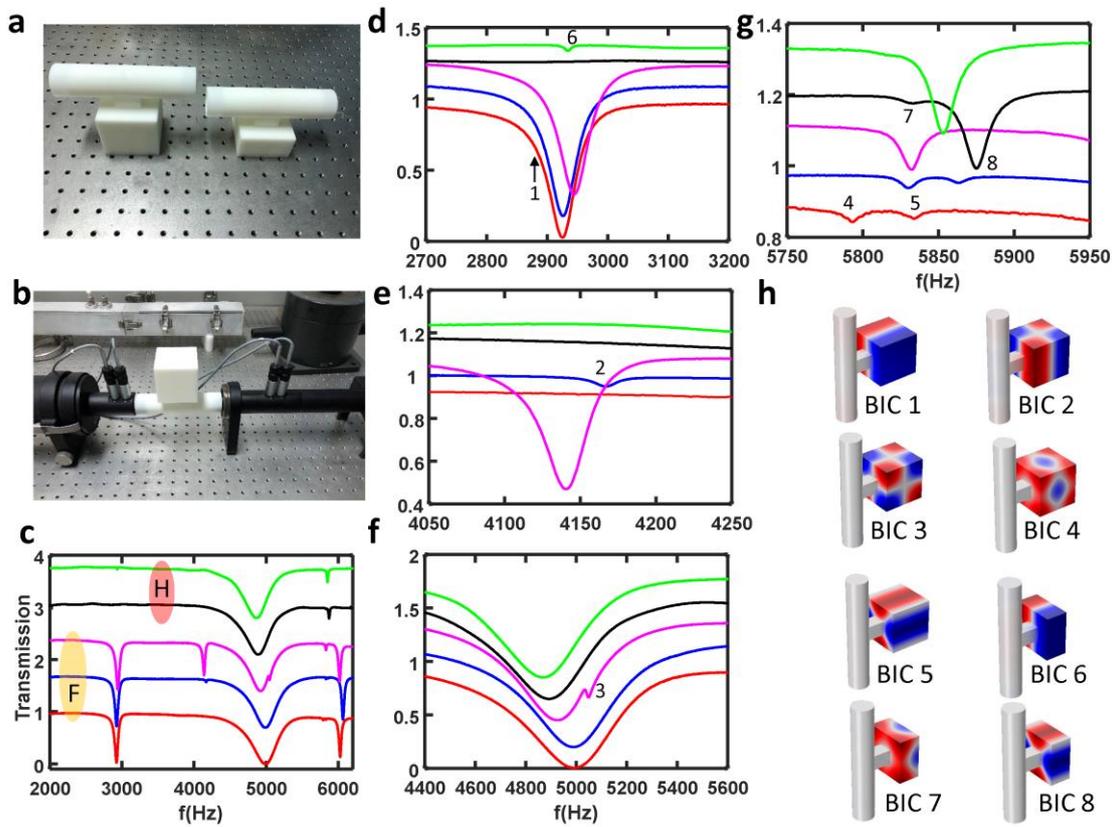

**Fig.3| Experimental verification in an open resonator**. **a,** Image of 3D printed full (left) and half (right) acoustic resonator. **b** Transmission measurement system. **c,** Full transmission spectrum in frequency range 2000-6000Hz. Red, blue and magenta lines represent full resonator (Lx=Ly=Lz=60mm) with yc=zc=0mm, yc=1 and zc=0mm, and yc=zc=4mm, respectively; Black and green lines represent half resonator (Lx=30mm, Ly=Lz=60mm) with yc=zc=0mm, and yc=1 and zc=0mm. Eight BICs are observed and labelled as 1-8. **d-g,** zoom in transmission spectrum in frequency range 2700-3200Hz (**d**),

4050-4250Hz (**e**), 4400-5600Hz (**f**), and 5750-5950Hz (**g**). **h,** pressure field distribution of eight BICs in the full and half resonator, which is also correlated to peaks 1-8 in the transmission spectrum.

To study the symmetry protected BIC, we fix Lx=Ly=Lz=60mm (Lx=30mm, Ly=Lz=60mm) for full (half) resonator and only change the centre shift $y_c$ and $z_c$ for the cuboid resonator with respect to the symmetric axis of the neck. From Fig.3d-f, it can be found that there are three symmetry protected BICs. For example, when $y_c$ increases from 0 to 1mm for the half resonator, a dip shows at around 2920Hz (green curve), which corresponds to mode $M_{121}$ (See Fig.3h-BIC 6). However, we did not observe $M_{121}$ (BIC 1) for the full resonator when the same asymmetry parameter yc=1mm is introduced. This can be explained by that this mode's Q-factor decreases relatively slowly with respect to $y_c$ and makes the resonance vanish in the low-Q mode background spectrum. Further enlarging $y_c$ may help to excite this mode.

Interestingly, another symmetry protected BIC $M_{221}$ (BIC 2) appears for a full resonator with yc=1(blue curve), evidenced by a shallow dip in the transmission spectrum (See Fig.3e). For mode $M_{222}$ (BIC 3), it is not enough to introduce the asymmetry by shifting $y_c$. The mode becomes visible in the transmission spectrum when $y_c$ and $z_c$ are adjusted to 4mm simultaneously, as shown in Fig.3f (magenta curve). We also systematically study the role of $y_c$ (or $y_c=z_c$) on the Q factor for BIC mode $M_{121}$, $M_{221}$ and $M_{222}$. The simulated and measured transmission spectra for the modes $M_{121}$ and $M_{221}$ are presented in Fig.4d-e and Fig.4g-h, respectively while the cases of $M_{222}$ can be found in Fig.S14. Good agreement can be found between these two. When $y_c$ is reduced to zero, the vanishing line width of resonances indicates the BIC's appearance. The Q-factor can be obtained by fitting the spectrum with the Fano formula[29] (See Section 2 in SI and Fig.S15). Fig.5a-b shows the measured Q-factor vs $y_c$ for the former two modes while the Q-factor of $M_{222}$ is put in Fig.S16. The maximum Q-factor for these two modes is only 250, lower than the theoretical prediction (see Fig.5e-f). This is because there are loss in the real system due to the thermo-viscous boundary layers, which degrades the Q-factor. Moreover, the Q-factor reduces with the increasing $y_c$, matching the trend of Q-factor vs $y_c$. Here, it is worth noting that the Q-factor for yc=0mm should be larger than 250. However, we cannot retrieve the exact value because they are almost indistinguishable from the background, which is exactly the signature of BIC.

For the two-mode interference induced BIC, we fix Lx=Ly=60mm and yc=zc=0mm, but only vary Lz. The size ratio Rx and Ry are defined as Rx=Lz/Lx and Ry=Lz/Ly, respectively. Thus, we have R=Rx=Ry for Lx=Ly. For the fixed Lz=60mm (Rx=Ry=1), we can find from Fig.3g that there are two BICs (BICs 4-5) in the range 5750-5950Hz, where the pressure field distributions are shown in Fig.3h (mode 4 and 5 for the full resonator, and mode 7-8 for the half resonator). Unlike symmetry-protected BICs, these two BICs always exist regardless of the value of $y_c$. Their Q-factors only depend on the size ratio. The simulated and measured transmission spectra for full resonator are presented in Fig.4f and Fig.4i where resonator height Lz varies from 59.5mm to 60.5mm. The measurement results of the half resonator with different Lz can be found in Fig.S17. Fig.4b-c shows the Q-factor of two modes as functions of size ratio. For the full resonator case, the maximum Q-factor is 583 at R=1 for mode B($M_{113}$) while it is 485 at R=1.008 for mode A ($M_{131}$). Note that the trend of measured Q factor vs R devitates from the theoretical prediction (See Fig.5e-f) because of inevitable intrinsic loss in the real system. The theoretical calculated Q-factor for mode A is maintained at a high value because the size ratio $R_{xy}$=Ly/Lx is always 1, which is precisely the critical size ratio. For mode B, the Q-factor reaches the maximum at critical size ratio Rx=Ry=1. However, the majority of Q-factors are ranged between 400-500. Such high-Q factors are already good enough for real applications, such as ultra-narrowband acoustic absorbers and enhanced acoustic emission. A similar phenomenon can also be observed for the half resonator. The only difference is that the maximum Q-factor for mode A (BIC 4) and B (BIC 5) are 393 and 327, respectively, which are lower than the full resonator case. The reduction of Q-factor may be attributed to increased thermo-viscous losses arising from the acoustic resonator's smaller volume.

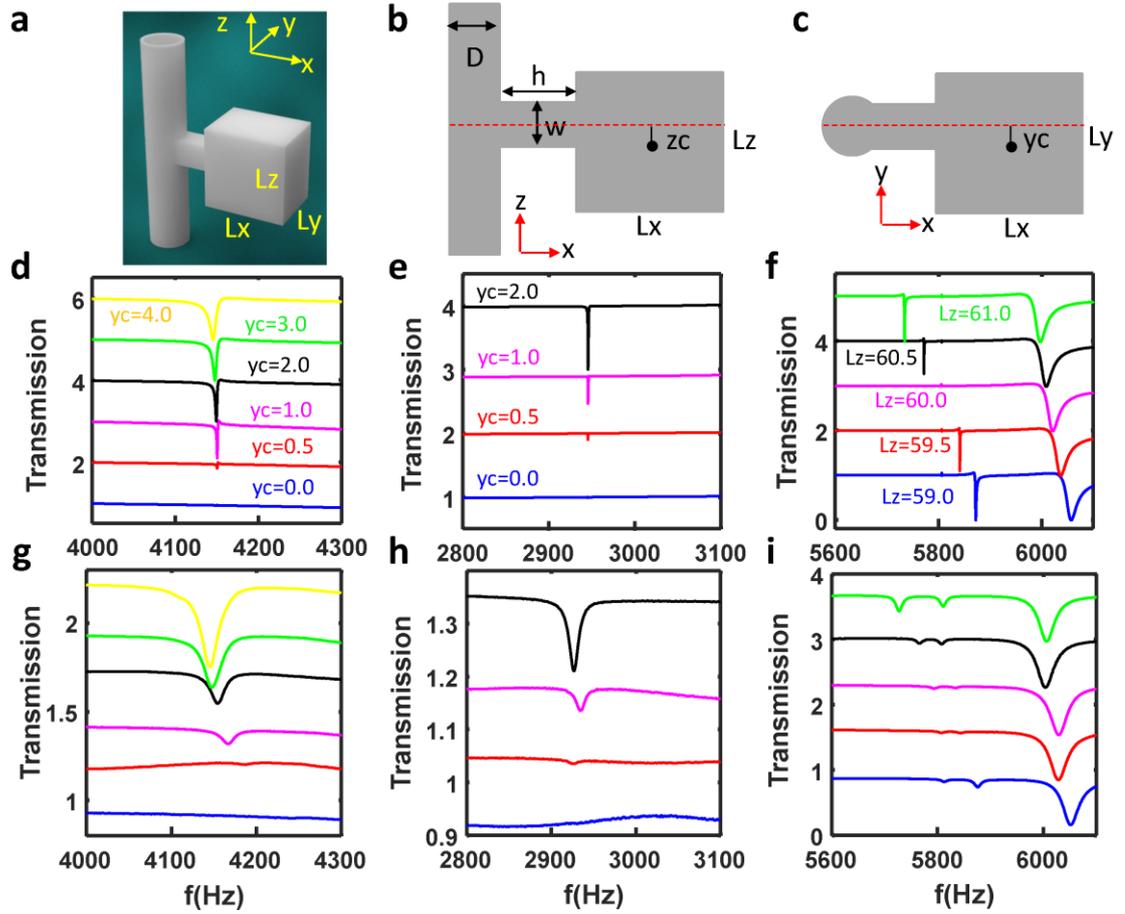

**Fig.4| Transmission spectra of an open resonator**. **a,** Schematic drawing of the 3D open resonator. **b,** Cross-section in XoZ plane for the 3D open resonator. **c,** Cross-section in XoY plane for the 3D open resonator. **d-f,** Simulated transmission spectra for full resonator (Lx=Ly=Lz=60mm) with different $y_c$ (**d**), half resonator (Lx=30mm, Ly=Lz=60mm) with different $y_c$ (**e**) and full resonator (Lx=Ly=60mm) for different values of Lz (**f**). **g-i,** Measured transmission spectra for full resonator (Lx=Ly=Lz=60mm) with different $y_c$ (**g**), half resonator (Lx=30mm, Ly=Lz=60mm) with different $y_c$ (**h**) and full resonator (Lx=Ly=60mm) for different values of Lz (**i**).

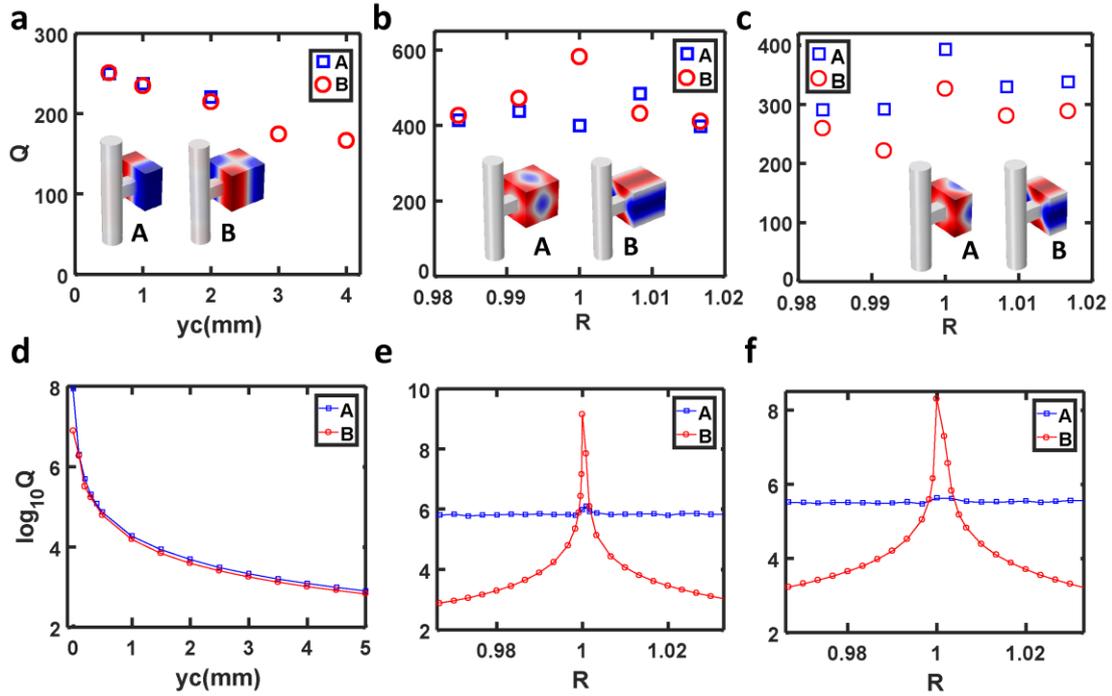

**Fig.5| Q-factor of QBIC**. **a,** Measured Q-factor of modes A ($M_{121}$) and B($M_{221}$) for the half resonator and full resonator as a function of $y_c$. **b,** Measured Q-factor of modes A ($M_{131}$) and B ($M_{113}$) for full resonator as a function of size ratio. **c,** Measured Q-factor of modes A ($M_{131}$) and B ($M_{113}$) for half resonator as a function of size ratio. **d,** Calculated Q-factor of modes A ($M_{121}$) and B($M_{221}$) for the half resonator and full resonator as a function of $y_c$. **e,** Calculated Q-factor of modes A ($M_{131}$) and B ($M_{113}$) for full resonator as a size ratio function. **f,** Calculated Q-factor of modes A ($M_{131}$) and B ($M_{113}$) for half resonator as a size ratio function.

## 3. Conclusion

We report the theoretical design of three types of acoustic BICs, including symmetry-protected BICs, modes interference induced BICs and mirror-symmetry induced BICs in a simple open resonator. Different types of BICs can be induced by tuning the resonators' geometrical parameters. Following the design principle, we fabricate such acoustic resonators and experimentally demonstrate these BICs by measuring the transmission spectrum. We found that the largest Q-factors retrieved from the transmission spectrum are 250, 583 and 393 for symmetry-protected QBIC, modes interference induced QBIC and mirror-symmetry induced QBIC. Such high-Q factors may bring more possibilities in designing acoustic devices with high performance, such as acoustic filters and acoustic emitters.

## Materials and Methods

### Simulations

All simulations in this paper are performed with the commercial software COMSOL Multiphysics. The speed of sound and air density is 349 m/s (corresponding to the experimental temperature of 30°C) and 1.29kg/m$^3$, respectively. When calculating the eigenmodes and transmission (or reflection spectrum), we apply perfect matched layer boundaries at the two ends of waveguides to mimic acoustic wave propagation in the infinite space. The other exterior boundaries are set as rigid.

### Experiments

The experimental samples are fabricated by 3D-printing technology using laser sintering stereolithography (SLA, 140μm) with a photosensitive resin (UV curable resin), exhibiting a manufacturing precision of 0.1 mm. The complex transmission (and reflection) coefficients of the samples are measured using a Brüel & Kjær type-4206T impedance tube with a diameter of 29 mm. A loudspeaker generates a plane wave, and the amplitude and phase of local pressure are measured by four 1/4-inch condenser microphones (Brüel & Kjær type-4187) situated at designated positions. The complex transmission (and reflection) coefficients are obtained by the transfer matrix method.

## Acknowledgements

L. Huang and A. E. Miroshnichenko were supported by the Australian Research Council Discovery Project (DP200101353) and the UNSW Scientia Fellowship program. Y. K. Chiang and D. A. Powell were supported by the Australian Research Council Discovery Project (DP200101708), S. Huang, Y. Cheng and Y. Li were supported by National Natural Science Foundation of China (Grants No. 12074286 and No. 11704284).

## Author Contributions

L. Huang, Y. K. Chiang and A. E. Miroshnichenko conceived the idea. L. Huang, Y. K. Chiang, and C. Shen, F. Deng performed the theoretical calculation and numerical simulation. S. Huang. Y. Cheng, B. Jia and Y. Li fabricated the sample and performed the reflection/transmission spectra measurements. L. Huang, Y. Li, D. A. Powell, and

# References

1. Ma, G. & Sheng, P. Acoustic metamaterials: From local resonances to broad horizons. *Sci. Adv.* **2**, e1501595 (2016).

2. Assouar, B. *et al.* Acoustic metasurfaces. *Nat. Rev. Mater.* **3**, 460–472 (2018).

3. Cummer, S. A., Christensen, J. & Alù, A. Controlling sound with acoustic metamaterials. *Nat. Rev. Mater.* **1**, 16001 (2016).

4. Ma, G., Yang, M., Xiao, S., Yang, Z. & Sheng, P. Acoustic metasurface with hybrid resonances. *Nat. Mater.* **13**, 873–878 (2014).

5. Li, Y. & Assouar, B. M. Acoustic metasurface-based perfect absorber with deep subwavelength thickness. *Appl. Phys. Lett.* **108**, 63502 (2016).

6. Li, Y. *et al.* Experimental Realization of Full Control of Reflected Waves with Subwavelength Acoustic Metasurfaces. *Phys. Rev. Appl.* **2**, 64002 (2014).

7. Hsu, C. W., Zhen, B., Stone, A. D., Joannopoulos, J. D. & Soljačić, M. Bound states in the continuum. *Nat. Rev. Mater.* **1**, 16048 (2016).

8. Hsu, C. W. *et al.* Observation of trapped light within the radiation continuum. *Nature* **499**, 188–191 (2013).

9. Plotnik, Y. *et al.* Experimental observation of optical bound states in the continuum. *Phys. Rev. Lett.* **107**, 28–31 (2011).

10. Marinica, D. C., Borisov, A. G. & Shabanov, S. V. Bound states in the continuum in photonics. *Phys. Rev. Lett.* **100**, 1–4 (2008).

11. Monticone, F. & Alù, A. Embedded Photonic Eigenvalues in 3D Nanostructures. *Phys. Rev. Lett.* **112**, 213903 (2014).

12. Molina, M. I., Miroshnichenko, A. E. & Kivshar, Y. S. Surface Bound States in the Continuum. *Phys. Rev. Lett.* **108**, 70401 (2012).

13. Weimann, S. *et al.* Compact Surface Fano States Embedded in the Continuum of Waveguide Arrays. *Phys. Rev. Lett.* **111**, 240403 (2013).

14. Xu, L. *et al.* Dynamic Nonlinear Image Tuning through Magnetic Dipole Quasi-BIC Ultrathin Resonators. *Adv. Sci.* **6**, (2019).

15. Kodigala, A. *et al.* Lasing action from photonic bound states in continuum. *Nature* **541**, 196 (2017).

16. Koshelev, K. *et al.* Subwavelength dielectric resonators for nonlinear nanophotonics. *Science (80-. ).* **367**, 288 LP-292 (2020).

17. Huang, C. *et al.* Ultrafast control of vortex microlasers. *Science (80-. ).* **367**, 1018 LP-1021 (2020).

18. Hein, S. & Koch, W. Acoustic resonances and trapped modes in pipes and tunnels. *J. Fluid Mech.* **605**, 401–428 (2008).

19. Hein, S., Koch, W. & Nannen, L. Trapped modes and Fano resonances in two-dimensional acoustical duct-cavity systems. *J. Fluid Mech.* **692**, 257–287 (2012).

20. Lyapina, A. A., Maksimov, D. N., Pilipchuk, A. S. & Sadreev, A. F. Bound states in the continuum in open acoustic resonators. *J. Fluid Mech.* **780**, 370–387 (2015).

21. Evans, D. V. & Linton, C. M. Trapped modes in open channels. *J. Fluid Mech.* **225**, 153–175 (1991).

22. Vassiliev, D. Existence Theorems for Trapped Modes. *J. Fluid Mech.* **261**, 21–31 (1994).

23. Davies, E. B. & Parnovski, L. Trapped modes in acoustic waveguides. *Q. J. Mech. Appl. Math.* **51**, (1998).

24. Huang, S. *et al.* Extreme Sound Confinement From Quasibound States in the Continuum. *Phys. Rev. Appl.* **14**, 21001 (2020).

25. Fan, S., Suh, W. & Joannopoulos, J. D. Temporal coupled-mode theory for the Fano resonance in optical resonators. *J. Opt. Soc. Am. A* **20**, 569–572 (2003).

26. Koshelev, K., Lepeshov, S., Liu, M., Bogdanov, A. & Kivshar, Y. Asymmetric Metasurfaces with High-$Q$ Resonances Governed by Bound States in the Continuum. *Phys. Rev. Lett.* **121**, 193903 (2018).

27. Friedrich, H. & Wintgen, D. Interfering resonances and bound states in the continuum. *Phys. Rev. A* **32**, 3231–3242 (1985).

28. Landi, M., Zhao, J., Prather, W. E., Wu, Y. & Zhang, L. Acoustic Purcell Effect for Enhanced Emission. *Phys. Rev. Lett.* **120**, 114301 (2018).

29. Miroshnichenko, A. E., Flach, S. & Kivshar, Y. S. Fano resonances in nanoscale structures. *Rev. Mod. Phys.* **82**, 2257–2298 (2010).